\documentclass[prl,aps,twocolumn,superscriptaddress,showpacs,amsmath,amssymb,floatfix]{revtex4}

\usepackage{graphicx}

\newcommand{\be}{\begin{equation}}
\newcommand{\ee}{\end{equation}}

\begin{document}
\title{Attractive Fermi gases with unequal spin populations in highly elongated traps}

\author{G. Orso}
\affiliation{BEC-INFM and Dipartimento di Fisica, Universita' di Trento, 
1-38050 Povo, Italy}
\affiliation{LENS and Dipartimento di Fisica, Universita' di Firenze, via N. Carrara 1, 50019 Sesto Fiorentino, Italy}

\begin{abstract}
We investigate two-component attractive 
Fermi gases with imbalanced spin populations in trapped one dimensional configurations.
The ground state properties are determined within local density approximation, starting 
from the exact Bethe-ansatz equations for the homogeneous case.
We predict that the atoms are distributed according to a {\sl two-shell} structure: a 
partially polarized phase in the center of the trap  and either 
a fully paired or a fully polarized phase in the wings. The partially polarized core is expected
to be a superfluid of the FFLO type. 
The size of the cloud as well as the critical spin polarization needed to suppress the fully paired shell, 
are calculated as a function of the coupling strength. 
\end{abstract}
\maketitle

Recent experimental studies  of polarized Fermi gases 
near a Feshbach resonance have attracted a great deal of interest \cite{ketterle}\cite{rice}. 
Since a finite spin polarization destabilizes the Bardeen-Cooper-Schrieffer (BCS) s-wave superconductivity, 
these strongly interacting systems are indeed promising candidates
to search for exotic pairing mechanisms.

So far, the experiments have mainly explored three-dimensional (3D) trapped configurations.
The MIT data \cite{ketterle} show that in the presence of a {\sl shallow} trap the polarized Fermi gas
phase-separates into {\sl three} concentric shells:  a fully paired superfluid 
core, an intermediate partially polarized 
normal shell and, finally, a fully polarized phase at the edge of the cloud. 
The Rice experiments have instead been performed on more anisotropic cigar-shaped
configurations and the obtained measurements suggest a somewhat different scenario \cite{rice}.  
A plausible explanation is that in the Rice experiments the applied trapping potential in
the {\sl radial} directions is tight enough that the gas cannot be considered as locally 
homogeneous and qualitatively new effects are seen which cannot be accounted for by the 
3D local density approximation. 

In this Letter, we study the limiting case where the radial confinement is so strong that 
only the lowest transverse mode is populated and the gas is kinematically 
one-dimensional (1D). We show that the ground state of the 1D polarized Fermi gas in a shallow axial trap
exhibits a {\sl two}-shell structure, in sharp contrast with the 3D case.
Our analysis is based on the {\sl exact} equation of state for a 1D homogeneous Fermi gas with 
arbitrary strong attractive interactions and the 1D local density approximation. 

Strongly interacting 1D Fermi gases have already been investigated experimentally
using ultracold atoms in optical lattices \cite{esslinger}. The possibility to create spin imbalanced
configurations opens new prospects to detect the Fulde-Ferrel-Larkin-Ovchinnikov (FFLO) state
\cite{fulde} \cite{note} under clean experimental conditions. In particular,
it was recently argued by Yang \cite{kunyang} that 
the ground state of 1D homogeneous attractive Fermi gases with unequal spin populations 
is the 1D analogue of the FFLO state. For trapped configurations, this implies that the 
{\sl partially polarized} phase of the 1D gas is a superfluid of the FFLO type  
and the experimentally measurable cloud sizes and density profiles provide indirect 
signatures of this unusual state. 
This is clearly exciting, because FFLO superconductivity  
has been of interest to condensed matter physicists for decades and, more recently, it has attracted 
the attention of the particle physics community \cite{casalbuoni}.


Consider a two-component atomic Fermi gas in the presence of an axially symmetric  
harmonic potential
\be
V_{trap}({\bf r})=\frac{1}{2}m\omega_\perp^2 r_\perp^2+\frac{1}{2}m\omega_z^2 z^2,
\ee 
where $m$ is the atom mass and $r_\perp=\sqrt{x^2+y^2}$. We require the 
anisotropy parameter $\lambda=\omega_z/\omega_\perp$ to be sufficiently small
that only the lowest transverse mode is populated. At zero temperature, this condition implies that
the Fermi energy $\epsilon_{F\uparrow}$ associated with the longitudinal motion of the majority $\uparrow$
component is much smaller than the distance between
energy levels in the transverse direction,  $\epsilon_ {F\uparrow}\ll \hbar \omega_\perp$, or, equivalently,
 $\lambda \ll 1/N_\uparrow$. Under this assumption, 
Bergeman et al. \cite{olshani} have shown that the scattering properties of atoms 
are well described by an effective 1D contact interaction $g_{1D}\delta(x)$, with
\begin{equation}
g_{1D}=\frac{2\hbar^2a_{3D}}{m a_\perp^2} \frac{1}{1-A a_{3D}/a_\perp} \;,
\label{g1d}
\end{equation}
where $a_{3D}$ is the 3D scattering length, $a_\perp=\sqrt{\hbar/m\omega_\perp}$ is the transverse 
oscillator length and  $A= 1.0326$. The effective 1D interaction
is attractive, $g_{1D}<0$, if $1/a_{3D}<A/a_\perp$. By tuning the 3D scattering length
near a Feshbach resonance, 
the coupling constant (\ref{g1d}) can be changed adiabatically 
from the weakly interacting regime, $g_{1D}\rightarrow 0^-$, to the strong coupling limit,   
$g_{1D}\rightarrow -\infty$.

The effective 1D Hamiltonian $H_{1D}$ acting on the fermions is given by 
$H_{1D}=H_{1D}^0+\sum_{i=1}^N V_{ho}(z_i)$, where $V_{ho}=m\omega_z^2 z^2/2$ is the longitudinal trapping
potential and
\be\label{hamiltonian}
H_{1D}^0=-\frac{\hbar^2}{2m}\sum_{i=1}^N\frac{\partial^2}{\partial z_i^2} + g_{1D}\sum_{i=1}^{N_\uparrow}
\sum_{j=1}^{N_\downarrow}\delta(z_i-z_j),
\ee
with $N=N_{\uparrow}+N_{\downarrow}$ being the total number of atoms. 
The Hamiltonian (\ref{hamiltonian}) is exactly soluble by Bethe's ansatz \cite{gaudin}.

Unpolarized 1D Fermi gases, in both homogeneous and trapped configurations, have been
investigated in Ref.\cite{tokatly} and Ref.\cite{alessio}, as an exactly solvable model for the   
BEC-BCS crossover. The homogeneous case with a finite spin polarization has been recently studied in 
Ref.\cite{batchLONG}.
In the following we investigate the magnetic properties of spin polarized 1D 
attractive Fermi gases in harmonic traps at zero temperature. 
We first focus on the homogeneous case by setting  $V_{ho}=0$. The effects of the axial confinement 
will be discussed later, using the 1D local density approximation (LDA). 

For fixed values of the linear number densities 
$n_\uparrow=N_\uparrow/L$ and $n_\downarrow=N_\downarrow/L$, where $L$ is the size of the system, 
the ground state energy $E$ of $H_{1D}^0$ is given by \cite{gaudin} 
\be\label{energy}
\frac{E}{L}=\frac{4 \hbar^2}{m a_{1D}^3} \left[\int_{-B}^{B}
(2\lambda^2 - \frac{1}{2})\sigma(\lambda)d\lambda
+\int_{-Q}^{Q}k^2 \rho(k)dk \right].
\ee
The spectral functions $\sigma(\lambda),\rho(k)$ in Eq.(\ref{energy}) are solutions of the
coupled integral equations \cite{gaudin}
\begin{eqnarray}
 &\sigma(\lambda)&= \frac{1}{\pi}-\int_{-B}^{B}
K_1(\lambda,\lambda^\prime)\sigma(\lambda^\prime) d\lambda^\prime-\int_{-Q}^{Q} K_2(\lambda,k)\rho(k) dk\nonumber, \nonumber\\
&\rho(k)&=\frac{1}{2\pi}-\int_{-B}^{B}K_2(\lambda,k)\sigma(\lambda) d\lambda,\label{gaudin}
\end{eqnarray} 
where the kernels are given by $K_1(\lambda,\lambda^\prime)=1/[\pi+\pi (\lambda^\prime-\lambda)^2]$ and
$K_2(\lambda,k)=2/[\pi+4\pi (k-\lambda)^2]$. Here $B$ and $Q$ are non-negative numbers related to 
the number densities by  
$n_\downarrow a_{1D}=2\int_{-B}^B \sigma(\lambda) d\lambda$ and $n_\uparrow a_{1D}-
n_\downarrow a_{1D}=2\int_{-Q}^Q\rho(k) dk $, respectively, and $a_{1D}=-2\hbar^2/mg_{1D}$ is the effective 1D scattering length. 
The coupling strength is controlled by the parameter $na_{1D}$, where $n=n_\uparrow+n_\downarrow$ is the total density:
the  weakly interacting case corresponds to $na_{1D}\gg 1$, whereas the strongly attractive regime is achieved for 
$na_{1D}\ll 1$.

The leading terms in the weak coupling expansion for the energy can be calculated by perturbation theory, yielding
$E/L=\hbar^2 \pi^2(n_\downarrow^3+n_\uparrow^3)/6m-2\hbar^2 n_\uparrow n_\downarrow 
/(m a_{1D})$. In the strongly interacting regime, corresponding to small values of $B$ and $Q$, we expand 
the spectral functions $\sigma(\lambda)$ and $\rho(k)$ around $\lambda=0$ and $k=0$, 
respectively, retaining up to quadratic terms. The coefficients of the expansion can be easily obtained from
Eq.(\ref{gaudin}). From Eq.(\ref{energy}), we then obtain $E=E^0+\Delta E$, where the leading term is given by
\be\label{tonks}
\frac{E^0}{L}=-\epsilon_B n_\downarrow+\frac{\hbar^2 \pi^2}{12m} n_\downarrow^3+
\frac{\hbar^2 \pi^2}{6m} (n_\uparrow-n_\downarrow)^3
\ee
and $\Delta E=\hbar^2 \pi^2 a_{1D}
[n_\downarrow^3(2n_\uparrow-n_\downarrow)+8n_\downarrow(n_\uparrow-n_\downarrow)^3]/12  m$.
The above result is in agreement with Ref.\cite{batchLONG}. Equation (\ref{tonks}) shows that in the 
strongly interacting regime the system is a mixture of a Tonks gas of $N_\downarrow$ 
diatomic molecules of mass $2m$ and binding energy $\epsilon_B=\hbar^2/ma_{1D}^2$, and an ideal 
Fermi gas formed by the unpaired  $\uparrow$ fermions.


For a fixed value of the total density $n\equiv n_{\uparrow}+n_{\downarrow}$,
the density difference $s\equiv n_{\uparrow}-n_{\downarrow}$ can vary in the range $0\leq s\leq n$.
For $s=0$, the ground state of $H_{1D}^0$ is {\sl fully paired} with a gap in the spin 
excitation spectrum \cite{alessio} whereas in the opposite limit $s=n$, 
the system is a {\sl fully polarized} gas of $\uparrow$ fermions.  
For $0<s<n$, the gas is {\sl partially polarized} and it is expected to be 
a superfluid of the FFLO type \cite{kunyang}. 

From Eq.(\ref{energy}), we calculate the chemical potential $\mu \equiv \partial (E/L)/\partial n$
and the {\sl effective} magnetic field $h\equiv\partial (E/L)/\partial s$, which are related to 
the chemical potentials
$\mu_\sigma$ of the two components by $\mu =(\mu_{\uparrow}+\mu_{\downarrow})/2$ and 
$h=(\mu_{\uparrow}-\mu_{\downarrow})/2$. From Eq.(\ref{energy}) it follows that
\begin{eqnarray}
&\mu&=\epsilon_B F_\mu(\tilde n,\tilde s),\nonumber\\
&h&=\epsilon_B F_h(\tilde n,\tilde s),\label{transf}
\end{eqnarray}
where $F_\mu$ and $F_h$ are universal functions of the dimensionless number densities $\tilde n\equiv n a_{1D}$ and
$\tilde s\equiv s a_{1D}$.
The energetic stability of the mixture
in the partially polarized phase allows us to use $\mu$ and $h$ as new coordinates, since 
the Jacobian of the transformation (\ref{transf}) is positive \cite{pethick}. 
 
\begin{figure}[tb]
\begin{center}
\includegraphics[width=6.8cm,angle=270]{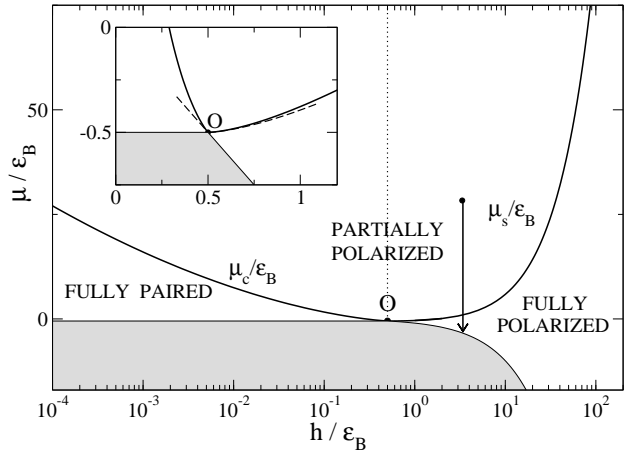}
\caption{Ground state magnetic properties for the homogeneous system. The dotted vertical line corresponds
to $h=\epsilon_B/2$. The arrow shows the trajectory of the local chemical potential in a trap.
Inset: zoom of the phase diagram near the point O=$(\epsilon_B/2,-\epsilon_B/2)$: the asymptotic 
behaviors are shown with dashed lines.}
\label{qpd}
\end{center}
\end{figure}

By setting $s=0$ in Eq.(\ref{transf}) we obtain the critical curve $\mu_c/\epsilon_B$ 
shown in Fig.\ref{qpd} and
corresponding to the phase boundary with the fully paired phase. We see that $\mu_c$ diverges logarithmically for vanishing 
magnetic fields and {\sl decreases} as $h$ increases, reaching its minimum value $\mu_c=-\epsilon_B/2$ 
for $h\rightarrow \epsilon_B/2^-$. The corresponding asymptotic behavior $\mu_c \simeq -h$ 
is plotted with a dashed line in the inset of Fig.\ref{qpd}, where we have magnified 
the region of the phase diagram near the point O=$(\epsilon_B/2,-\epsilon_B/2)$.
Similarly to what happens for the attractive Hubbard model \cite{woynarovich}, the critical 
magnetic field $h_c= h(n,0)$ is related to the energy gap $\Delta$ in the spin sector for the unpolarized case 
by $h_c=\Delta/2$. Since $\Delta$ measures the strength of interactions, $h_c$ increases by decreasing the density
$n$ (or, equivalently, the chemical potential $\mu$) reaching $h_c=\epsilon_B/2$ in the strongly attractive limit
$\mu\rightarrow -\epsilon_B/2$, as shown in Fig.\ref{qpd}.

By setting $s=n$ in Eq.(\ref{transf}) we obtain the second critical curve $\mu_s/\epsilon_B$ plotted in Fig.\ref{qpd} and
corresponding to the phase boundary with the fully polarized phase.
In this particular case, the saturation field $h(n,s=n)$ and the corresponding chemical potential 
$\mu_s=\mu(n,s=n)$ can be calculated in closed form from Eqs (\ref{energy}) and (\ref{gaudin}), yielding 
\be\label{hs}
h=2\epsilon_B[Q^2(1-\arctan(2Q)/\pi)+(2Q-\arctan(2Q)+\pi)/4\pi]
\ee
and $\mu_s=-h+2Q^2\epsilon_B$, where $Q=\pi na_{1D}/2$. In the weak coupling limit, 
corresponding to $h\gg \epsilon_B$, the chemical potential diverges as 
$\mu_s=h-4\sqrt{h \epsilon_B}/\pi+(4/\pi^2-1/2)\epsilon_B$ whereas in the strong coupling regime 
it approaches O as
$\mu_s=-\epsilon_B/2+(2\sqrt{2}/3\pi\sqrt{\epsilon_B})(h-\epsilon_B/2)^{3/2}$.
This latter behavior is also shown in the inset of Fig.\ref{qpd} with dashed line.

In Fig.\ref{qpd}, the fully paired and the fully polarized phases are limited from
below by a shaded area representing the vacuum state, i.e. the state of zero total density.
For $h<\epsilon_B/2$, the minimum value of the chemical potential $\mu\equiv \mu_v$ is given by 
$\mu_v=-\epsilon_B/2$, whereas  for $h>\epsilon_B/2$, we find $\mu_v=-h$, since 
$\mu+h=\mu_\uparrow$ is the chemical potential of the majority component.

We see from Fig.\ref{qpd} that for a fixed value of the magnetic field, only two phases are
allowed for the 1D system: partially polarized and fully paired if $h<\epsilon_B/2$ 
{\sl or} partially polarized and fully polarized if $h>\epsilon_B/2$. 
This is clearly different from the 3D case, where $\mu_c$ is an increasing function of the magnetic field 
\cite{mueller} and thus all three phases are allowed for a given $h$.

The effects of a shallow harmonic potential $V_{ho}(z)$ can be taken into account via LDA, which
is applicable provided the size of the cloud is large compared to the axial oscillator 
length $a_z=\sqrt{\hbar/m\omega_z}$, implying $N \gg 1$.
Under this assumption, the density profiles $n_{\sigma}(r)$ for the two components 
$\sigma=\uparrow, \downarrow$ are obtained by imposing the local equilibrium condition
$\mu_{\sigma}[n_\uparrow(z),n_\downarrow (z)]+V_{ho}(z)=\mu_{\sigma}^0$, where 
$\mu_{\sigma}[n_\uparrow,n_\downarrow]$ 
are the corresponding chemical potentials of the homogeneous system and $\mu_{\sigma}^0$ are constants 
fixed by the normalization $N_{\sigma}=\int n_{\sigma}(z) dz$. This is equivalent to
\begin{eqnarray}
&\mu[n_\uparrow (z),n_\downarrow (z)]&=\mu^0-V_{ho}(z),\nonumber\\
&h[n_\uparrow (z),n_\downarrow (z)]&=h^0, \label{LDA}
\end{eqnarray}
where $\mu^0, h^0=(\mu_\uparrow^0 \pm \mu_\downarrow^0)/2$.
Equation (\ref{LDA}) shows that the {\sl local} mean chemical potential decreases
by increasing the distance from the center of the trap while 
the {\sl local} magnetic field remains constant 
throughout. In the phase diagram of Fig.\ref{qpd}, this corresponds to follow  the
vertical line $h=h^0$ downwards starting from the point $\mu=\mu^0$, as shown by the arrow.

We immediately see that trapped gases with spin imbalanced components 
have a {\sl two-shell} structure, with an inner partially polarized phase 
in the center of the trap. 
The outer shell is {\sl either} fully paired, if $h^0<\epsilon_B/2$, {\sl or} fully polarized, 
if $h^0>\epsilon_B/2$.
Hence, the fully paired and the fully polarized phases are {\sl mutually exclusive} in a trap.
For the special case $h^0=\epsilon_B/2$, shown with dotted line in Fig.\ref{qpd},  the partially 
polarized phase extends to the edge of the cloud. 

In order to relate $\mu^0,h^0$ to the total number $N \equiv N_\uparrow+N_\downarrow$ of atoms
and the spin polarization $P\equiv(N_\uparrow - N_\downarrow)/N$, we need to 
calculate the density profiles, 
starting from Eqs (\ref{energy}) and (\ref{LDA}). The normalization conditions
can be expressed in terms of the dimensionless chemical potential $\tilde \mu^0=\mu^0/\epsilon_B$
and the dimensionless densities $\tilde n$ and $\tilde s$, as
\begin{eqnarray}
N \frac{a_{1D}^2}{a_z^2}=\sqrt{2}\int_0^{\tilde \mu_v} 
\frac{\tilde n(\tilde \mu^0-x)}{\sqrt{x}}dx, \nonumber \\
(N_\uparrow - N_\downarrow)\frac{a_{1D}^2}{a_z^2}=\sqrt{2} \int_0^{\tilde \mu_v}
\frac{\tilde s(\tilde \mu^0-x)}{\sqrt{x}}dx, \label{scaletrap}
\end{eqnarray}
where ${\tilde \mu_v}=\mu_v/\epsilon_B$.
Equation (\ref{scaletrap}) shows that, in the presence of the trap, the coupling strength is given by $Na_{1D}^2/a_z^2$,
which is large in the weakly interacting
regime and vanishes in the strongly attractive limit.

\begin{figure}[tb]
\begin{center}
\includegraphics[width=6.7cm,angle=270]{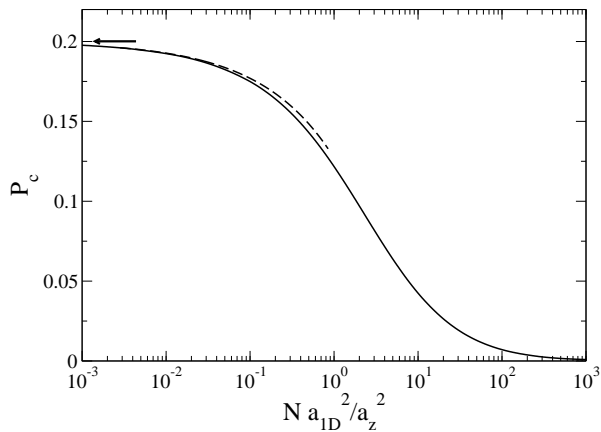}
\caption{Critical spin polarization as a function of the interaction 
parameter $N a_{1D}^2/a_z^2$. The behavior in the strong coupling regime is 
shown with dashed line. The arrow corresponds to the asymptotic limit $P_c=0.2$.}
\label{figPc}
\end{center}
\end{figure}

The condition $h^0=\epsilon_B/2$ yields the critical spin polarization $P=P_c$ needed to suppress 
the fully paired 
shell; for $P>P_c$ the outer shell is always fully polarized. In Fig.\ref{figPc} we plot the 
critical spin polarization as a function of the parameter $Na_{1D}^2/a_z^2$ (solid line). We see that $P_c$ 
increases going towards the strong coupling regime where it saturates to $P_c=0.2$. This value 
is surprisingly small compared to the critical polarization $P_c^{3D}\sim 0.77$ 
found \cite{ketterle} in 3D Fermi gases at resonance $(a_{3D}\rightarrow\infty)$. 
The corresponding density profiles in the strongly attractive limit, where Eq.(\ref{tonks}) holds,
are given by $n_\downarrow(z)=(8m/\pi^2\hbar^2)^{1/2}  (\mu^0+\epsilon_B/2-V_{ho}(z))^{1/2}$,
with $\mu^0+\epsilon_B/2=\hbar \omega_z N_\downarrow/2$, and $n_\uparrow(z)=n_\downarrow(z)+
(2m/\pi^2\hbar^2)^{1/2} (\mu^0 + h^0-V_{ho}(z))^{1/2}$, with 
$\mu^0 + h^0=(N_\uparrow - N_\downarrow)\hbar \omega_z$.
The inclusion of the next order correction $\Delta E$ yields  
$P_c=1/5(1-0.364 \sqrt{Na_{1D}^2/a_z^2})$. This asymptotic behavior is shown in 
Fig.\ref{figPc} with dashed line.

\begin{figure}[tb]
\begin{center}
\includegraphics[width=6.7cm,angle=270]{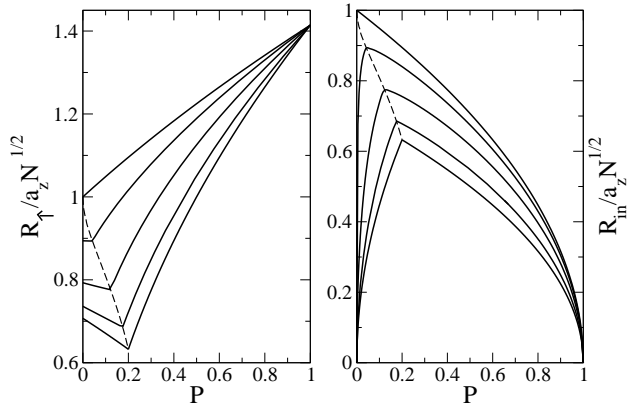}
\caption{Size of the cloud (left) and radius of the inner shell (right) 
as a function of the spin polarization for different values of the parameter 
$N a_{1D}^2/a_z^2=\infty,10,1,0.1,0$. The non monotonic behavior is a signature of the two-shell 
structure. }
\label{radii}
\end{center}
\end{figure}

For $P\neq P_c$, the cloud exhibits a two-shell structure, with  
the radius $R_{in}$ of the inner shell fixed by the condition
$\mu^0-V_{ho}(R_{in})=\mu_{c}(h^0)$, for $h^0<\epsilon_B/2$, and $\mu^0-V_{ho}(R_{in})=\mu_{s}(h^0)$, 
otherwise.
The size $R_\uparrow$ of the cloud, where the total density vanishes, is instead given by the condition
$\mu^0-V_{ho}(R_\uparrow)=\mu_v(h^0)$.
Both quantities are plotted in Fig.\ref{radii} 
as a function of the spin polarization $P$ for different values of the parameter $Na_{1D}^2/a_z^2$.
In the absence of interaction (top curve), we find $R_\uparrow=\sqrt{1+P}a_zN^{1/2}$
and $R_{in}=\sqrt{1-P}a_zN^{1/2}$, showing that $R_\uparrow(R_{in})$ increases (decreases)
as $P$ increases. 
For finite interactions, we see that $R_\uparrow$ {\sl decreases} as $P$ increases until $P=P_c$, 
and it {\sl increases} for larger values of the spin polarization. In particular, in the 
strong coupling regime, we find   
$R_{\uparrow}=\sqrt{(1-P)/2}a_z N^{1/2}$ for $P\leq 1/5$, and $R_{\uparrow}=\sqrt{2P}a_z N^{1/2}$ for $P>1/5$.
This non monotonic behavior is a signature of  the outer shell. Indeed, an increase in the 
spin polarization implies
that the radius of the density profile for the minority component decreases, as $N_\downarrow$ decreases.    
For $P<P_c$, the profiles for the two components match in the fully paired outer shell, and thus also $R_\uparrow$ decreases.

The inner radius (right panel) shows the opposite behavior: $R_{in}=0$ for zero spin polarization 
 and it increases rapidly until $P=P_c$, where $R_{in}=R_\uparrow$ and the fully paired shell disappears. 
For higher values of the polarization, $R_{in}$ decreases and finally vanishes at $P=1$. 
In particular, in the strong 
coupling regime, we find $R_{in}=\sqrt{2P}a_z N^{1/2}$, for $P\leq 1/5$, and 
$R_{in}=\sqrt{(1-P)/2}a_z N^{1/2}$ for $P>1/5$.

The above scenario for a 1D polarized Fermi gas can be investigated in current experiments
with ultracold atoms under a tight radial confinement, e.g. $^{40}$K samples in an array of 1D tubes 
created with optical lattices \cite{esslinger}. In order to reach a strongly degenerate regime,
the thermal energy $k_B T$ should be small compared to the Fermi energy, implying 
$k_B T \ll N \hbar \omega_z$. This condition can be satisfied by taking the  
typical experimental parameters \cite{esslinger}
$\omega_z/2\pi \sim 250$ Hz for the axial frequency, mean number of particles (per tube) $N \sim 100$ 
and $T \sim 50$nK.  
The spin imbalance can be easily induced by a radio-frequency sweep and the density profiles
measured by absorption imaging techniques.

In conclusion, we have presented an exact solution of an experimentally accessible 1D model 
of a polarized Fermi gas at arbitrary strong coupling. We have found that, 
differently from the 3D case, 
the trapped gas phase separates in a two-shell structure, with a FFLO superfluid core surrounded by 
either a fully paired or a fully polarized phase, depending on the value of the
spin polarization. 
Experimental signatures of this novel and exotic state are a reduced value of the 
critical spin polarization $P_c$ and a non monotonic behavior of the size of the cloud as a 
function of the spin polarization.

We acknowledge fruitful discussions with N. Prokof'ev, S. Giorgini, W. Zwerger and G.V. Shlyapnikov. 
We are also grateful to Z. Idziaszek and G. Astrakharchik for useful suggestions on numerics.
Finally, we want to thank G.V. Shlyapnikov for the kind hospitality at LPTMS (Orsay) during the 
preparation of this manuscript.
This work was supported by the Ministero dell'Istruzione, dell'Universita' e della Ricerca (M.I.U.R.).



\begin{thebibliography}{16}
\bibitem{note} The distinctive feature
of the FFLO state  is that the unpaired fermions induce an oscillatory behavior in the 
superconducting correlation function
$\langle \Psi(z)_\uparrow^\dagger \Psi(z)_\downarrow^\dagger \Psi_\downarrow(z^\prime)
\Psi_\uparrow (z^\prime)\rangle$ 
in the large distance limit $|z-z^\prime|\rightarrow \infty $.
\end{thebibliography}
\end{document}